\documentclass[
aps, pra,
amsmath,amssymb,
11pt,
final,
tightenlines,
twoside,
twocolumn,
nofloats,
nofootinbib,
superscriptaddress,
showkeys,
showkeywords,
]
{revtex4-2}

\usepackage[T2A]{fontenc}
\usepackage[utf8x]{inputenc}
\usepackage[russian,english]{babel}
\usepackage{graphicx}
\usepackage{dcolumn}
\usepackage{bm}
\usepackage{amsmath}
\usepackage{siunitx}

\input{maik.rty}

\setcitestyle{authoryear,round}
\def\squareforqed{\hbox{\rlap{$\sqcap$}$\sqcup$}}

\def\sq{\ifmmode\squareforqed\else{\unskip\nobreak\hfil
\penalty50\hskip1em\null\nobreak\hfil\squareforqed
\parfillskip=0pt\finalhyphendemerits=0\endgraf}\fi}

\def\utw{\smash{\rlap{\lower5pt\hbox{$\sim$}}}}

\def\udtw{\smash{\rlap{\lower6pt\hbox{$\approx$}}}}

\def\diameter{{\ifmmode\mathchoice
{\ooalign{\hfil\hbox{$\displaystyle/$}\hfil\crcr
{\hbox{$\displaystyle\mathchar"20D$}}}}
{\ooalign{\hfil\hbox{$\textstyle/$}\hfil\crcr
{\hbox{$\textstyle\mathchar"20D$}}}}
{\ooalign{\hfil\hbox{$\scriptstyle/$}\hfil\crcr
{\hbox{$\scriptstyle\mathchar"20D$}}}}
{\ooalign{\hfil\hbox{$\scriptscriptstyle/$}\hfil\crcr
{\hbox{$\scriptscriptstyle\mathchar"20D$}}}}
\else{\ooalign{\hfil/\hfil\crcr\mathhexbox20D}}%
\fi}}

\begin{document}
	
	\selectlanguage{english}
	
	
	
	\title{OPTICAL AND X-RAY VARIABILITY OF $\gamma$~CAS STARS: SAO\,49725 }
	
	\author{\firstname{A.~F.}~\surname{Kholtygin}}
	\email{afkholtygin@gmail.com}
	\affiliation{St. Petersburg State University, St. Petersburg, Russia}
	
	\author{\firstname{I.~Ya.}~\surname{Yakunin}}
	\affiliation{St. Petersburg State University, St. Petersburg, Russia}
	\affiliation{Special Astrophysical Observatory RAS, Russia}
	
	\author{\firstname{V.~S.}~\surname{Bukharinov}}
	\affiliation{St. Petersburg State University, St. Petersburg, Russia}
	
	\author{\firstname{D.~N.}~\surname{Mokshin}}
	\affiliation{St. Petersburg State University, St. Petersburg, Russia}
	
	\author{\firstname{E.~B.}~\surname{Ryspaeva}}
	\affiliation{Crimean Astrophysical Observatory RAS, Russia}
	
	\author{\firstname{O.~A.}~\surname{Tsiopa}}
	\affiliation{Main (Pulkovo) Astronomical Observatory RAS, Russia}
	
	\begin{abstract}
		The present paper is devoted to studying the variability of the $\gamma$~Cas type star SAO\,49725. Both the optical and X-ray spectra of the star are analyzed. Variability of line profiles in the spectra of SAO\,49725 was discovered on short (70-223 minutes) scales. Based on the TESS photometric light curves of SAO\,49725  the regular variations of light curves 
		are detected with a period of 1.1989 days identified as the rotation period of the star. The pattern of photometric variability in SAO 49725, as observed by TESS, varies significantly across different epochs. The TESS components of the SAO\,49725 light curves with periods of $\sim3-21\,$days may be instrumental.
		
	\end{abstract}
	
	\maketitle
	
	
	\section{INTRODUCTION}
	
	An important subclass of main sequence (MS) stars of spectral type B are Be stars with emission lines in the spectrum and rotation velocities close to critical (classical Be stars). A feature of this group of stars is the occurrence of a circumstellar disk, the presence of which is identified by the Balmer spectral line emissions of hydrogen and, often, by the emission lines of other elements.
	
	These emission lines in the Be-star spectra are often two-component. At that, their shape is determined by the disk structure, the physical conditions in it and the angle at which the disk is inclined to the observer. The optical and X-ray spectra and the luminosities of Be-stars show variability on short time scales, while their disks are unstable, as evidenced by rapid variations in the emission line profiles up to their almost complete disappearance. Most of the Be stars are bright X-ray sources with a typical X-ray luminosity of $L_\mathrm{X}\sim10^{29}\,\mbox{эрг}/\mbox{с}$ erg s\(^{-1}\).
	
	Strong magnetic fields of up to tens of kilogauss have been detected in approximately 10\% of B stars. Many models of Be-star disk formation require a magnetic field \citep[see, e.g.,][]{Brown-2008}. At the same time, despite the decades of polarization observations of Be stars, no magnetic field has ever been detected in any of them, with the possible exception of the star $\lambda\,$Eri~\citep{Hubrig-2017}, suggesting that all Be stars are non-magnetic~\citep{GrunhutWade2012,Wade-2016}. According to~\cite{ud-Doula-2018}, even the presence of a weak magnetic field of 10-100 G in a rapidly rotating B star would lead to a rapid disk destruction.
	
	All the above-mentioned features of Be stars are identified for $\gamma$~Cas variables, a distinguished group of Be stars with X-ray luminosities by 1-2 orders of magnitude higher than those of normal Be stars, and an anomalously high ($kT~\sim\! 5-30\,$keV) plasma temperature, emitting in the X-ray spectral region. At the same time, the optical spectra of $\gamma$~Cas variables are not remarkable among similar spectra of other Be stars.
	
	The reasons for the formation of unusual X-ray emission in $\gamma$~Cas variables are still unclear. A comparison of the variability of their optical and X-ray spectra can shed light on the nature of $\gamma$~Cas stars. In this paper, such a comparison is made for the bright star SAO 49725 (BD+47 3129, RX J2030.5+4751, MWC 1023, TIC 187940144).
	
	The paper is organized as follows. Section~\ref{s.object} presents the data on the physical parameters of the star. Information on the analyzed optical and X-ray spectra of the star is given in Section~\ref{s.observ}. Section~\ref{s.OptVary} presents the spectroscopic, and Section ~\ref{s.X-rayPhotom} the photometric variability of the star according to the Transiting Exoplanet Survey Satellite (TESS) data. The study of the X-ray variability of SAO 49725 is presented in Section~\ref{s.X-rayPhotom}. The conclusions to the paper are presented in Section~\ref{s.Concl}.
	
	\begin{figure*}[]
		\centering
		\includegraphics[scale=1.30]{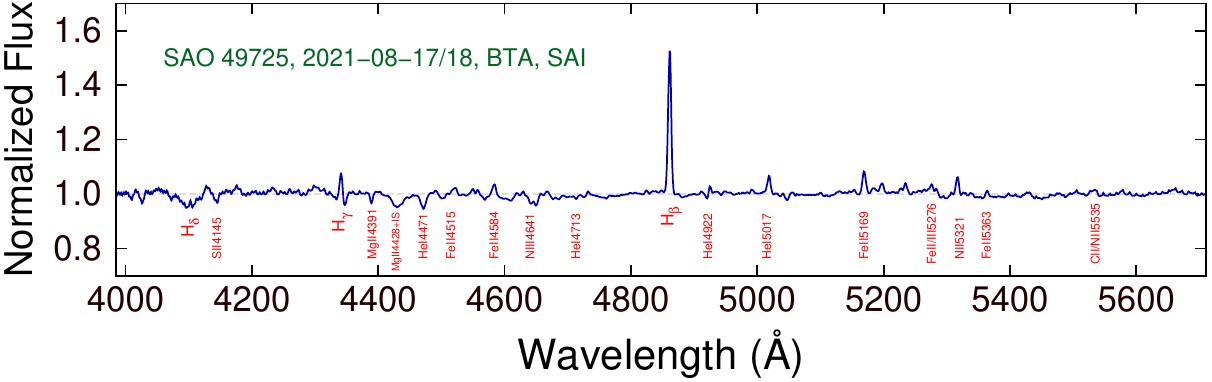}   
		\caption{The average normalized spectrum of SAO 49725 obtained with the SAO RAS BTA telescope on August 17/18, 2021. 
		}
		\label{Fig.AtlasSpectra_SAO49725}
	\end{figure*}
	
	\section{THE MAIN INFORMATION ABOUT THE STAR}
	\label{s.object}
	
	The star SAO\,49725 is classified as a $\gamma$~Cas variable object based on the analysis of its X-ray observations \citep{Smith-2016}. It is a binary system consisting of a massive Be star with a mass of about 13$M_{\odot}$ and a low-mass companion with a mass of $0.2-0.5\,M_\odot$ and an orbital period of $P_{\text{orb}} = 26^{\mathrm{d}}.11$ ~\citep{Naze-2022}. The radial velocity curve also admits a solution with an orbital period of 137 days and a companion mass of $0.5-0.7\,M_\odot$. In this paper, we analyze the spectra of the massive Be component of SAO\,49725.
	
	
	\begin{table}[ht!]
		\caption{Parameters of the star SAO\,49725}
		\label{tab:star_pars}
		\bigskip
		{\footnotesize
		\begin{tabular}{c|c|c}
			\hline
			Parameter & Value & Refs \\
			\hline
			Spectral type & B0.5 III--Ve & [1] \\
			$m_v$, mag & 9.27 & [1] \\
			$T_{\text{eff}} \times 10^{-3}$, K & 33.9 & [2] \\
			$R_*$, $R_\odot$ & 7.17 & [2] \\
			$\log g$ & 4.35 & [2] \\
			$\log (L_*/L_\odot)$ & 4.78 & [2] \\
			$M$, $M_\odot$ & 13 & [3] \\
			$d$, pc & 2384 & [4] \\
			$v \sin i$, km s$^{-1}$ & $182 \pm 33$ & [5] \\
			$i$, $\deg$ & $37 \pm 8$ & [2] \\
			$P_{\text{rot}}$, days & 1.20 & [2] \\
			$P_{\text{orb}}$, days & 26.11, 137.0 & [3] \\
			$M_{\text{comp}}$, $M_\odot$ & 0.2--0.5, 0.5--0.7 & [3] \\
			\hline
		\end{tabular}} \\
		\medskip
			\scriptsize{
				[1]\,\cite{Smith-2016}, [2]\,the present paper, [3]\,\cite{Naze-2022}, [4]\,Gaia EDR3, [5]\,\cite{Lopes_de_Oliveira-2006}.}
	\end{table}
		 
	
	Table~\ref{tab:star_pars} lists the parameters of the Be component in the SAO 49725 system.. The rotation period, $P_\mathrm{rot}=1^\mathrm{d}.20\,$, was found from the analysis of the TESS satellite observations (see Section~\ref{s.TessPhotom}).
	
	The luminosity $L_*$, effective temperature $T_\mathrm{eff}$ and acceleration of gravity $\lg g$ of the star were estimated from its magnitudes in the $ U $, $ B $ and $ V $ bands, using the relations presented by~\cite{Nieva2013}. The correction $A_V$ for interstellar extinction in the \( V \) filter was calculated using the \texttt{dustmap} library ({\sf dustmap}~\cite{Green2018}). The distance to the star was determined from its parallax adopted from the~\cite{GaiaEDR3-2020}. We estimated the radius of the star by the Stefan– Boltzmann formula using the $L_*$ and $T_\mathrm{eff}$ values we found.
	

	\section{OBSERVATIONS AND SPECTRAL REDUCTION}
	\label{s.observ}
	
  	We have carried out the observations of SAO~49725, analyzed in this paper, at the SAO RAS 6-m BTA telescope within the framework of the program called the Rapid Variability of Line Profiles in the Spectra of OBA Stars and the Nature of their X-Ray Emission (responsible applicant---A.\,F.~Kholtygin, St.~Petersburg State University), using the SCORPIO multi-mode focal reducer \citep{AfanasievMoiseev2005} on August~17, 2021. We have obtained 431 spectra of the star with a 5-s exposure each.
  	

  	\begin{figure*}[ht!]
  		\centering
  		\includegraphics[scale=0.4]{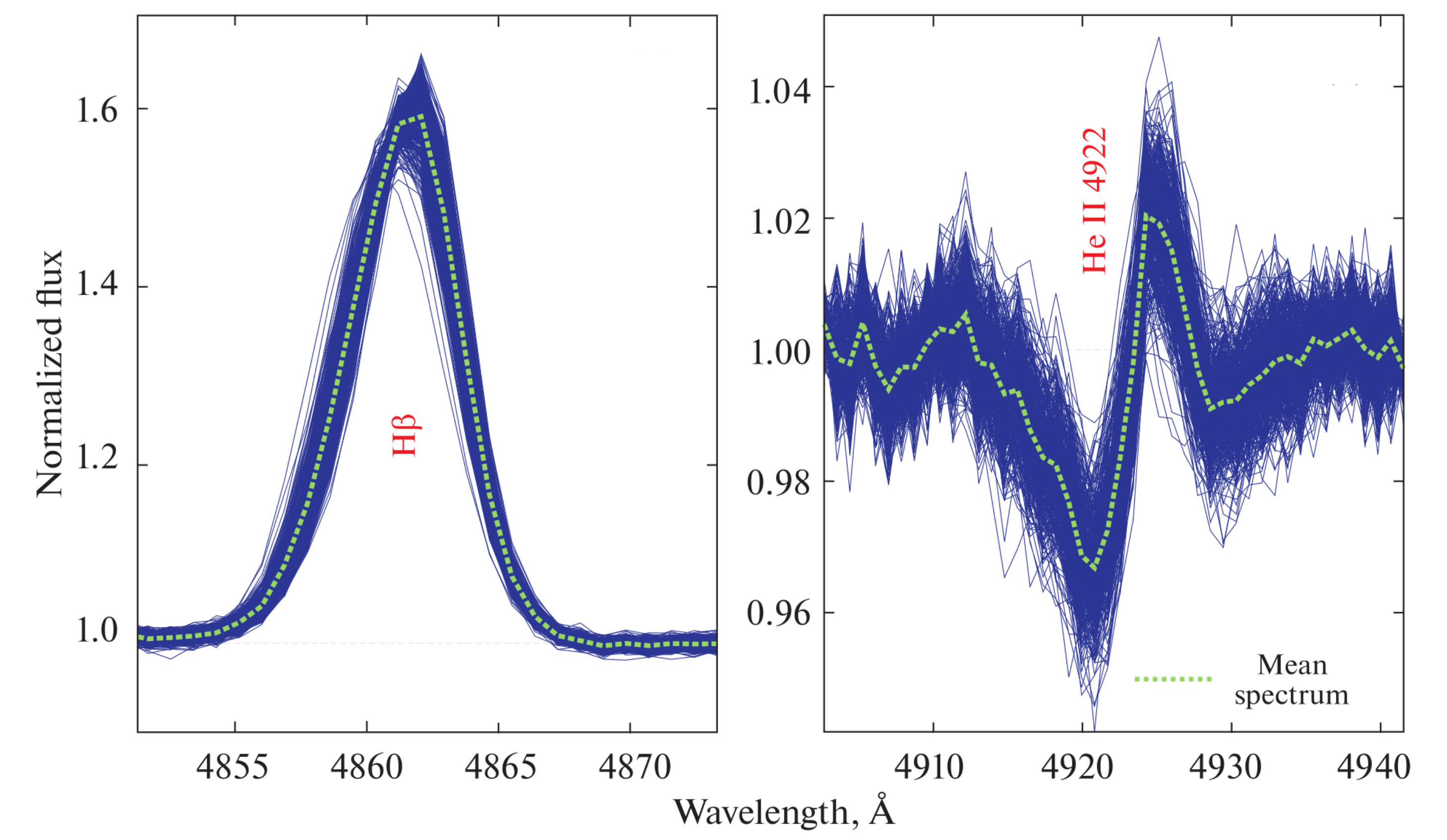}   
  		\caption{The proﬁles of H$\beta$ lines (left panel) and the He I $\lambda$ 4922 line (right panel) in the spectrum of SAO~49725 obtained with the BTA. Dashed lines depict the average proﬁles of the corresponding lines.}
  		\label{Fig.SAO49725_LPV_MeanSp}
  	\end{figure*}
  	
	Primary reduction and extraction of the spectra obtained with the BTA were performed in the MIDAS system. Standard procedures were used reducing the spectra: subtraction of the CCD chip bias, wavelength calibration, after which the one-dimensional spectra were reduced to the barycenter of the Solar System. During the wavelength calibration, the closest in time spectrum of the ThAr lamp was used for a series of spectra. All the BTA spectra we have obtained were normalized to the continuum. The normalization procedure is described in the paper by~\cite{Kholtygin-2006}. The average spectrum of the star SAO~49725, averaged over all obtained spectra and normalized to the continuum, is shown in Fig.~\ref{Fig.AtlasSpectra_SAO49725}.
	
	
	\section{OPTICAL VARIABILITY OF SAO\,49725}
	\label{s.OptVary}
	
	\subsection{Optical Variability}
	\label{ss.OpticalSpectra}
	
	Analysis of Fig.~\ref{Fig.AtlasSpectra_SAO49725} shows that the H$\beta$ line is a purely emission line, while the Balmer H$\gamma$ and H$\delta$ lines have a significant absorption component. Numerous Fe~II line profiles are also emissive. It can be assumed that the appearance of Fe~II lines in the emission is explained by the small depth of the absorption components of these lines in the spectrum of the star itself. The total line profile, which is the sum of the emission and absorption components, turns out to be emissive.
	
	As shown in Fig.~\ref{Fig.AtlasSpectra_SAO49725}, the He I line profiles exhibit both emission and absorption features. When analyzing line profile variations in the spectrum of a star, it is reasonable to choose lines of a sufficient depth and without a strong blending. Based on these principles, we selected the Balmer hydrogen lines, helium lines, and Fe~II ion lines.
	
	The variability of the H$\alpha$ and He~I $\lambda$4922 line profiles in the spectrum of SAO~49725 is illustrated in Fig.~\ref{Fig.SAO49725_LPV_MeanSp}. For each line, all 431 normalized profiles and the average line profile are shown. The shape of the H$\beta$ line profile corresponds to the orientation of the stellar disk at a small angle to the line of sight~\citep[see, for example,][]{Rivinius-2013}. The He~I $\lambda$~4922 line profiles have a typical shape of the P~Cyg-type profiles. The full width of the profile of this line corresponds to the expansion velocity of the stellar wind matter of about 800 km s$^{-1}$.
	
	\begin{figure*}[ht!]
		\includegraphics[scale=0.40]{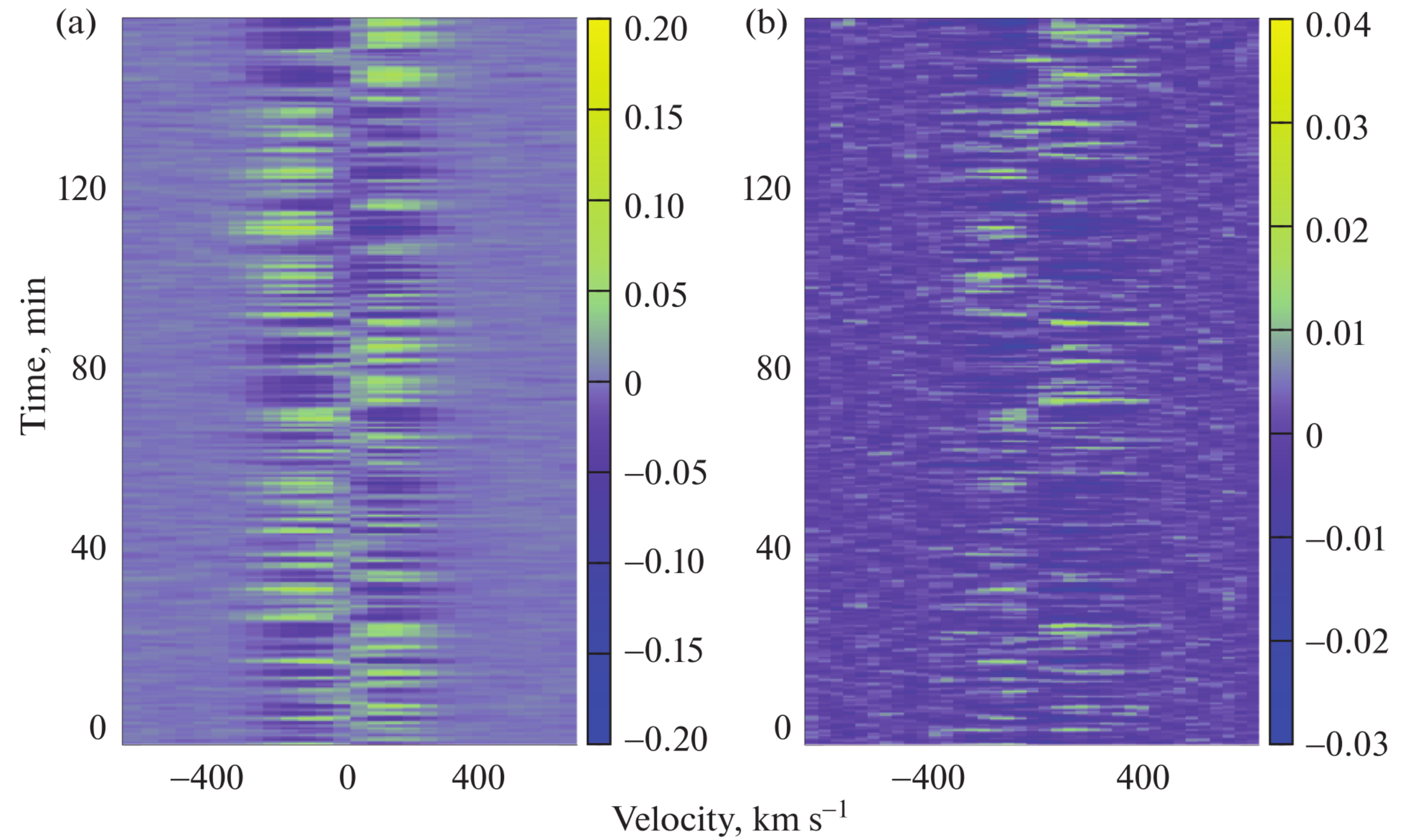}   
		\caption{Dynamic spectra of H$\beta$ (a) and He~I 4922 (b) line proﬁle variations.
		}
		\label{Fig.SAO49725_DiffSpMap}
	\end{figure*}
	
	For a visual representation of line profile variations, we will use their differential $d(\lambda)$ profiles:
	\begin{equation}
		\label{Eq.DiffProf}
		d(\lambda)=F_i(\lambda) - \overline{F}_i(\lambda)
	\end{equation} 
	where $F_i(\lambda)$ is the flux in the $i$-th spectrum, normalized to the continuum, and  $\overline{F}_i(\lambda)$ is the flux at wavelength $\lambda$ averaged over all observations.
	
	In the analysis of the differential profiles, instead of the wavelength $\lambda$, we will use the Doppler shifts (radial velocities) $V$ from the laboratory wavelength $\lambda_0$ of the line, $V = c (\lambda / \lambda_0 - 1)$, where $c$ is the speed of light. The amplitude of $H\beta$ line profile variations reaches 12\% in flux units in the continuum adjacent to the line, while the He~I and Fe~II line profile variations do not exceed 3-5\%. Figure~3 shows the dynamic spectra of the $H\beta$ and He~I $\lambda$4922 line profile variations, Figs.~\ref{Fig.SAO49725_DiffSpMap}a and \ref{Fig.SAO49725_DiffSpMap}b, respectively.
	
	\subsection{Regular Line Proﬁle Variability in the SAO\,49725 Spectrum}
	\label{ss.RegSpVary}
	
	To search for regular components in the line profile variations in the spectrum of SAO~49725, we performed a Fourier analysis of the difference line profiles $d(V)$. For each wavelength $\lambda_k$ on the line profile, corresponding to the Doppler shift $V = V_k$, the set of values $\{d(V_k, t_i)\}, i = 1, \ldots, N$, where $N$ is the number of analyzed profiles, represents the analyzed time series.
	
	The search for the periodic components of each of the series for the $V_k$ values within the line profile was carried out using the CLEAN method ~\cite{Roberts-1987}. In the Fourier spectrum (Schuster periodograms), regular components with the frequencies corresponding to the Fourier spectrum amplitude maxima, exceeding the value corresponding to the selected significance level $\alpha$ are selected.
	
	Table~\ref{Table.FourBTA} presents the frequencies (column~2) and periods (column~3) of possible harmonic components of the H, He~I and Fe~II line profile variations, found from the analysis of the SAO~49725 spectra we have obtained at the BTA. The significance levels of the components detected are given in the fourth column of Table~\ref{Table.FourBTA}.
	
	\begin{table}[ht]
		\centering
		\caption{\small Frequencies and periods of regular components of line proﬁle variations in the spectrum of SAO~49725 according to the BTA data}
		\label{Table.FourBTA}
		{\footnotesize \begin{tabular}{c|c|c|c|c|c|c}     \hline
				No. & $\nu,$, min$^{-1}$   & $P,$ min              &      $\alpha$      & ~H  &  HeI  & FeII    \\
				\hline  
				(1) & (2)    & (3) & (4)   & (5) & (6) & (7) \\  \hline 
				1  & 0.00447  & $223.61\pm319.47$ &  $10^{-3}$ & + & + & +   \\  
				2  & 0.00467  & $214.27\pm293.33$ &  $10^{-6}$ & + & + & +   \\  
				3  & 0.01436  & $ 69.61\pm30.96 $ &  $10^{-6}$ & + & + & +   \\  
				4  & 1.12769  & $  0.89\pm 0.01 $ &  $10^{-2}$ & + & - & +   \\  \hline
			\end{tabular}                                                    
		}                                                                                                                  
	\end{table}
	
	To estimate the error of the identified periods, the standard relationship $\Delta\nu = 1/T$ was used, according to, e.g.,~\cite{Vityazev-2001}, where $T = 156.52$ min is the total duration of the SAO~49725 observation time series at the BTA. The obtained frequencies and periods were not previously detected in the analysis of spectral and photometric variations of SAO~49725. The found periods are close to the typical periods of rapid line profile variations in the spectra of OBA stars (see, for example, ~\cite{Dushin-2013,Kholtygin-2018}).
	
	\section{TESS PHOTOMETRY OF SAO\,49725}
	\label{s.TessPhotom}
	
	\begin{figure*}[ht!]
		\includegraphics[scale=0.60]{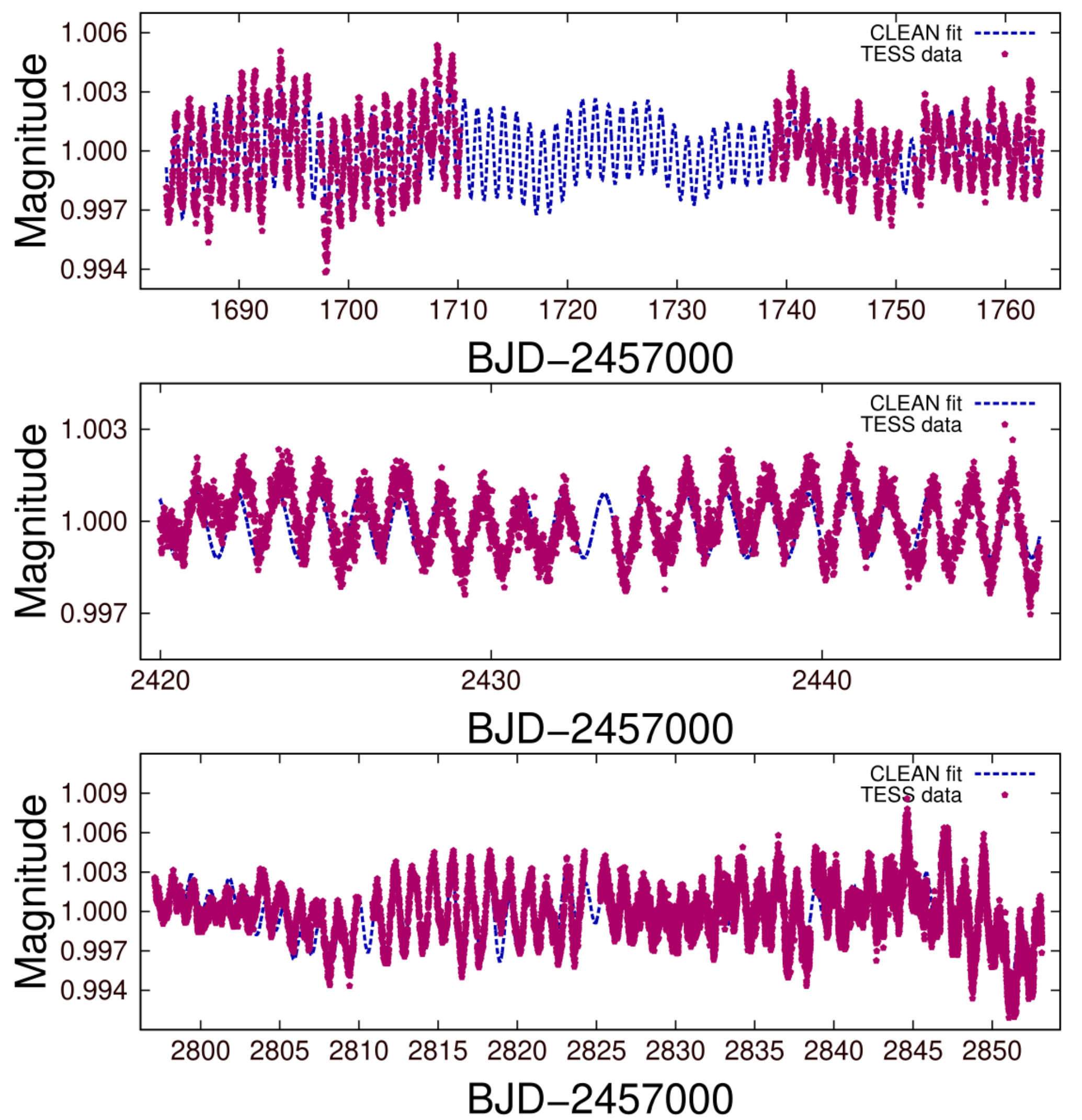}   
		\caption{The TESS SAO 49725 light curves in 2019 (Top panel), 2021 (Middle panel), and 2022 (Bottom panel). The dashed lines show the CLEAN light curve ﬁts.         
		}
		\label{Fig.TESS_SAO49725}
	\end{figure*}
	
	Photometric observations of OBA stars by the TESS satellite are currently available, described by~\cite{Jenkins-2016}. The TESS photometry data for SAO~49725 are available at  TASOC\footnote{TESS Data For Asteroseismology Lightcurves https://archive.stsci.edu/hlsp/tasoc}. SAO~49725 was observed by TESS in 2019, 2021 and 2022. TESS photometry in 2019 was carried out from July~18 to August~14 (28~days) and from September~12 to October~6 (25~days). In 2021---from July~24 to August~20 (28~days). In 2022---from August~5 to September~30 (57~days).
	
	The light curves of SAO~49725 retrieved from the TASOC website for the above epochs are shown in Fig.~\ref{Fig.TESS_SAO49725}. The minimum amplitudes of the star brightness variations were obtained in 2021. In 2021 and 2022, the brightness variation amplitudes exceed these values twice and thrice, respectively.
	
\begin{figure*}[ht!]
		\includegraphics[scale=1.060]{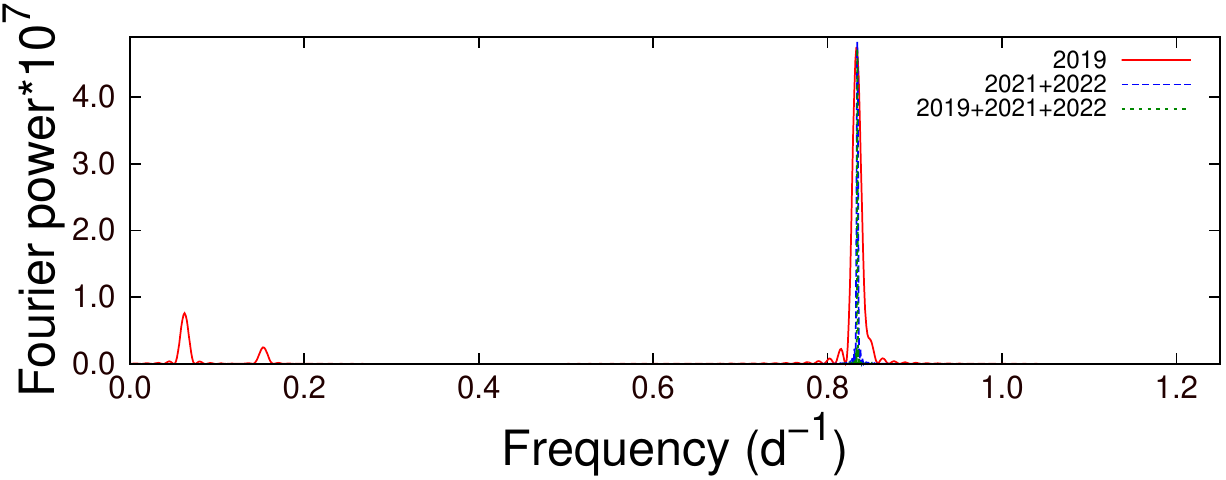}   
		\caption{The Fourier spectrum (CLEAN periodograms) of TESS brightness variations for the observations in 2019 (the solid curve), in 2021–2022 (the dashed line) and for the entire light curve (the dotted curve).         
		}
		\label{Fig.TESS_SAO49725FSp}
	\end{figure*}

	We have analyzed the photometric light curves in 2019, 2021, 2022 and the combined light curve for the entire observation period using the CLEAN method. Figure~\ref{Fig.TESS_SAO49725FSp} shows the corresponding Fourier periodograms. For ease of comparison, their amplitudes for the 2019 epoch are additionally multiplied by 0.55, and the periodograms of the 2021--2022 light curve---by 1.10. All the three periodograms coincide only for the frequency $\nu = 0.8341$\,d$^{-1}$, corresponding to the period of $P = 1^\mathrm{d}.1989 \pm 0^\mathrm{d}.0012$.
	
	\begin{table*}[ht]
		\centering
		\caption{\footnotesize Frequencies and periods of the periodic components of the SAO\,49725 brightness variations according to the TESS data.}
		\label{Table.FourierTESS}  
		{\footnotesize \begin{tabular}{c|c|c|c|c|c|c}     \hline
				~~~~No~~~~ & ~~~~$\nu,$ 1/d~~~~    & ~~~~~~~~$P,$ days~~~~~~~~ &  ~~~~2019~~~~    &~~~~2021-2022~~~~ & ~~~~All data~~~~ & ~~~~$\alpha$~~~~ \\  \hline
				1  & $0.0471$  & $21.23\pm1.041$   &  -   & +    & -   &$10^{-2}$  \\  
				2  & $0.0625$  & $15.99\pm9.53$    &  +   & -    & -   &$10^{-6}$  \\  
				3  & $0.1526$  & $ 6.55\pm1.60$    &  +   & -    & -   &$10^{-6}$  \\  
				4  & $0.3215$  & $ 3.11\pm0.36$    &  +   & -    & -   &$10^{-2}$  \\  
				5  & $0.8341$  & $1.1989\pm0.0012$ &  +   & +    & +   &$10^{-6}$  \\  \hline
			\end{tabular}                                                    
		}                                                                                                                  
	\end{table*}
	
	The most probable assumption is that the period $P = 1^\mathrm{d}.1989 \pm 0^\mathrm{d}.0012$ is the rotation period of the star. In support of this assumption, let us estimate the possible limiting values of the rotation period using the data from Table~\ref{tab:star_pars} and the standard relation:	
	\begin{equation}
		\label{Eq.ProtDays}
		P_\mathrm{rot}\, = \sin i\cdot \frac{50.61}{V\!\sin i\mbox{(km/s)}} \cdot\left(\frac{R_*}{R_{\odot}}\right) \quad [\mbox{days}].   
	\end{equation}
	where $v \sin i$ is measured in km s$^{-1}$, and $P_{\mathrm{rot}}$ in days. According to~\cite{Naze-2022}, the angle of inclination of the stellar rotation axis is in the range of $[30^\circ - 70^\circ]$, which corresponds to the rotation period values in the interval of $0.997\le P_\mathrm{rot}\le 1.873\,$, which includes the period we found, $P=1^\mathrm{d}.1989\pm0^\mathrm{d}.0012\,$. This value corresponds to the typical rotation periods of Be stars. Therefore, we can assume that the rotation period of SAO~49725 is
	$P_\mathrm{rot}=1^\mathrm{d}.1989\pm0^\mathrm{d}.0012\,$. Using this value and the rotation velocity $v \sin i$ values from Table~\ref{tab:star_pars} allows us to refine the rotation axis inclination angle: $i= 37^\circ\pm8^\circ$.
	
\begin{figure*}[ht!]
		\centering
		\includegraphics[scale=0.52]{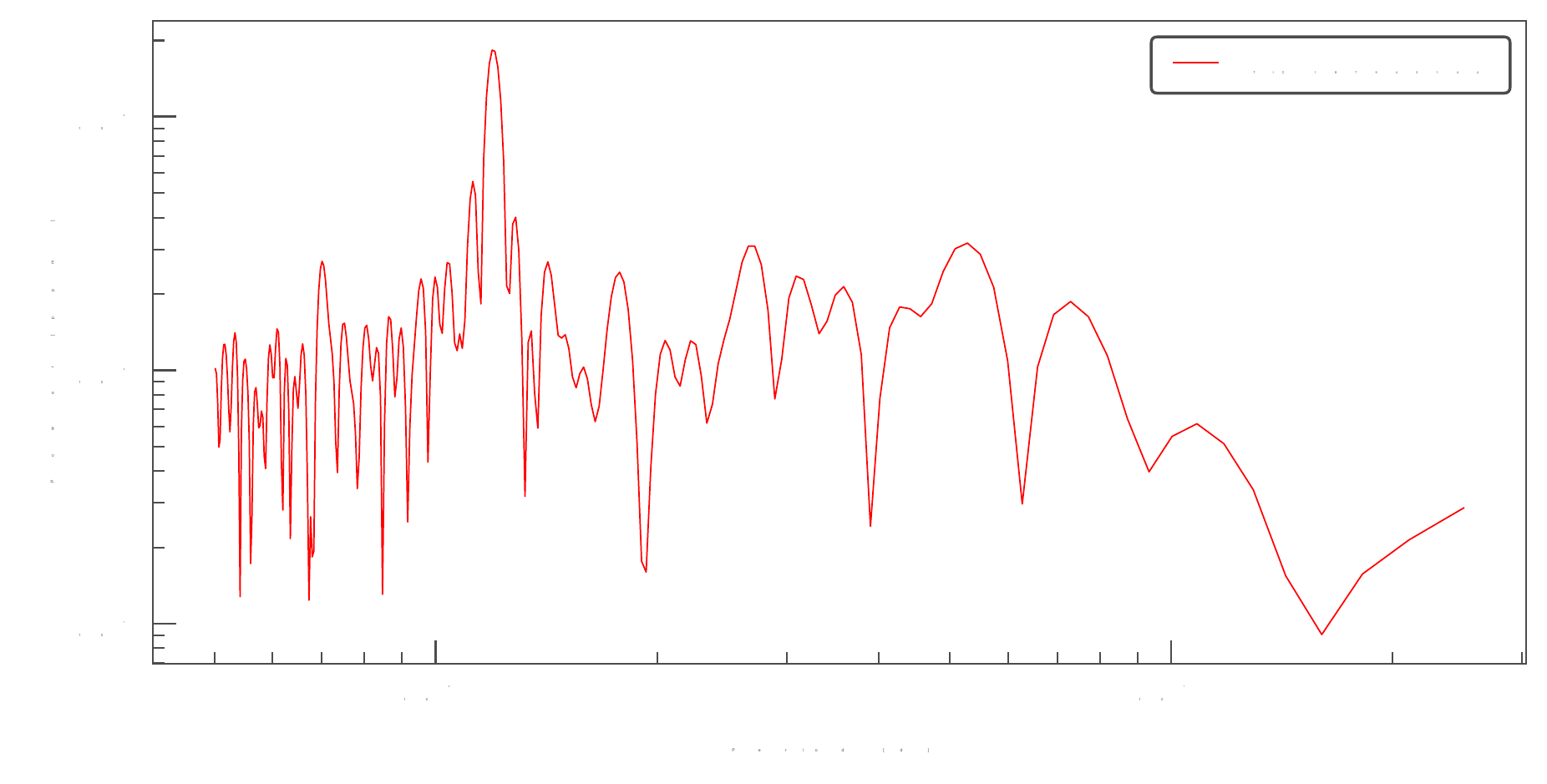}   
		\caption{The Fourier spectrum (a Lomble-Scargle periodogram) of the SAO\,49725 light curve based on the 2019 TESS observations (sector 14).
		}
		\label{Fig.X-raySAO49725-2019_LS}
	\end{figure*}

	The analysis of data in Table~\ref{Table.FourierTESS} shows that the light curve based on the 2019 TESS data contains, in addition to the main component corresponding to the rotation period of the star, the components $\nu_1 - \nu_3$ with the periods of $15^\mathrm{d}.99\pm9^\mathrm{d}.53\,$, $ 6^\mathrm{d}.55\pm1^\mathrm{d}.60\,$ and $3^\mathrm{d}.11\pm0^\mathrm{d}.36\,$. For the light curve of SAO~49725 obtained in 2019 (sector~14, observations from September~12 to October~6), we have also constructed the Lomb-Scargle periodogram \citep{Lomb1976,Scargle1982} shown in Fig.~\ref{Fig.X-raySAO49725-2019_LS}. This periodogram shows the same harmonic components as those presented in Table~\ref{Table.FourierTESS}.
	
	The nature of these components is still unclear. However, it can be noted that, according to ~\cite{Kholtygin-2023}, within the determination errors, they correspond to the periodic components obtained from the analysis of the light curve of HD~45995, a $\gamma$~Cas-type star according to the TESS data for 2018--2019: $14^\mathrm{d}.72\pm8^\mathrm{d}.72\,$, $6^\mathrm{d}.63\pm1^\mathrm{d}.86\,$ and $3^\mathrm{d}.91\pm0^\mathrm{d}.656$.
	
	The specified components were detected only in the observational epochs of 2018--2019 and are absent in the 2020-2022 epochs, which indicates their transient nature. The correspondence of the periodic components obtained in close observation epochs, but for different stars, indicates the instrumental nature of the brightness variations with the periods indicated above. The component with a period of $21^\mathrm{d}.23\pm1^\mathrm{d}.041\,$, recorded during the analysis of the TESS satellite observations in 2021--2022, may also be of an instrumental nature.
	
	\section{SAO 49725 IN XMM-NEWTON OBSERVATIONS}
	\label{s.X-rayPhotom}
	
	The analysis of X-ray spectra and X-ray light curves of early-type stars allows us to study the structure of their stellar winds and the formation of hot gas in them. The periods of variations of the optical line profiles in the spectra of a number of such stars correspond to the frequencies of their X-ray brightness variations \citep[see, for example,][]{Kholtygin-2022}.
	
	We have analyzed the archival X-ray observations of SAO~49725, carried out at the XMM-Newton observatory on December~9, 2003 (ObsID = 201200201). The procedures of primary data processing are described by~\cite{RyspaevaKholtygin2021}. In the paper cited we have presented the results of study of the SAO~49725 spectra. This paper deals with the stellar light curves' analysis and the results are given below.

	\subsection{Analysis of X-ray Light Curves}
	\label{ss.X-rayPhotomXMM}
	
	\begin{figure}[ht!]
		\centering
		\includegraphics[scale=0.60]{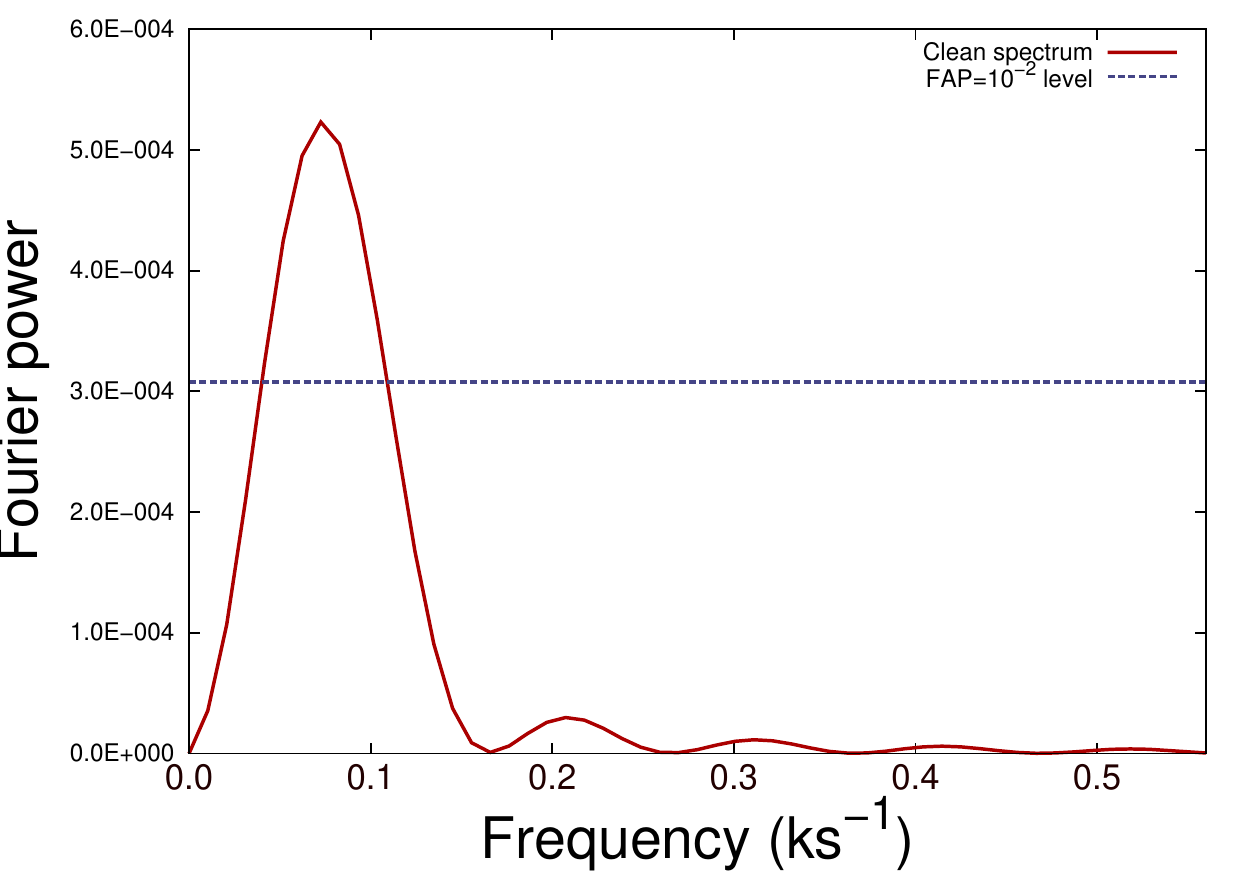}   
		\caption{The Fourier spectrum of the X-ray light curve of SAO 49725 based on the observations of December 9, 2003 with the EPIC camera of the XMM-Newton telescope in the energy ranges of 2-10\,keV.
		}
		\label{Fig.X-raySAO49725_FourSp}
	\end{figure}
	
	The total duration of observations of SAO~49725 was \SI{11515}{\second} (\SI{192}{\minute}). Analysis of the light curve by the CLEAN method showed the presence of a single regular component with a period of $P=230\pm178\,$min for data obtained with the PN detector at a significance level of $\alpha = 0.01$ for the observation band of \SIrange{0.2}{10}{\keV} (see Fig.~\ref{Fig.X-raySAO49725_FourSp}).
	
	Given that the identified component was detected only at a modest significance level, and its period is longer than the observation period, its reality should be confirmed by longer X-ray observations. Note that the third harmonic of this period $P/3=76.6\pm59.3\,$min is close to the period $P_3=69.61\pm30.96\,$min of line profile variations in the spectrum of SAO~49725 (see Table~\ref{Table.FourBTA}).
	
	\section{CONCLUSION}
	\label{s.Concl}
	
	This paper presents the results of a study of the optical and X-ray variability of the $\gamma$~Cas variable star SAO~49725. Based on the analysis of the optical and X-ray spectra and light curves of the star, the following conclusions can be drawn:
	
	\begin{itemize}
		\item The line profiles in the spectrum of SAO 49725 vary on short (70–223 minute) time intervals. The presence of ultrafast spectral variations with a period of $0.89 \pm 0.01$ minutes is possible.
		
		\item Analysis of the SAO~49725 light curves obtained by the TESS satellite allowed to determine the rotation period of the star ($P = 1^{\mathrm{d}}.1989 \pm 0^{\mathrm{d}}.0012$) and refine the inclination angle of its rotation axis ($i = 37^\circ \pm 8^\circ$).
		
		\item The light curves of SAO~49725 based on the 2019 TESS data contain components with the periods of about 3--21 days, the nature of which is still unclear, but may be instrumental.
	\end{itemize}
	
	\section*{ACKNOWLEDGMENTS}
	
	The observations conducted with the SAO RAS telescopes were supported by the Ministry of Science and Higher Education of the Russian Federation. The renovation of telescope equipment is currently provided within the national project ``Science and Universities''.
	
	\section*{FUNDING}
	
	A.\,F.\,Kh., I.\,A.\,Ya., D.\,N.\,M., and E.\,B.\,R. are grateful to the Russian Science Foundation for the financial support through the Russian Science Foundation grant No.~23-22-00090.
	
	\section*{CONFLICT OF INTEREST}
	
	The authors of this work declare that they have no conflicts of interest.
	
	\bibliographystyle{aspb1}                                                                                                                       
	\bibliography{Kholtygin-2024}

	\selectlanguage{english}
\end{document}